\documentclass[pra,twocolumn]{revtex4}
\usepackage{graphicx,amsmath,amssymb, bm, epstopdf, color}

\usepackage{hyperref}
\hypersetup{pdfborder = {0 0 0}}

\newcommand{\bp}{\bm\rho}
\newcommand{\sT}{\mathcal{T}}

\newcommand{\avg}[1]{\langle #1 \rangle}
\newcommand{\E}{\hat{E}}

\begin{document}
\title{Reflective Ghost Imaging through Turbulence}

\author{Nicholas D. Hardy}
\author{Jeffrey H. Shapiro}%
\affiliation{%
Research Laboratory of Electronics, Massachusetts Institute of Technology, Cambridge, Massachusetts 02139, USA
}%

\date{\today}

\begin{abstract}
Recent work has indicated that ghost imaging may have applications in standoff sensing. However, most theoretical work has addressed transmission-based ghost imaging. To be a viable remote-sensing system, the ghost imager needs to image rough-surfaced targets in reflection through long, turbulent optical paths. We develop, within a Gaussian-state framework, expressions for the spatial resolution, image contrast, and signal-to-noise ratio of such a system. We consider rough-surfaced targets that create fully developed speckle in their returns, and Kolmogorov-spectrum turbulence that is uniformly distributed along all propagation paths. We address both classical and nonclassical optical sources, as well as a computational ghost imager.
\end{abstract}


\maketitle

\section{Introduction}
Ghost imaging is a transverse active-imaging technique that exploits the correlation between two light beams to image a target without spatially-resolving measurements of the light beam that has undergone target interaction.  The two beams are the signal, which interacts with the target and then is measured by a single-pixel bucket detector, and the reference, which is directly measured by a high spatial-resolution detector. An intensity cross-correlation between the signal beam that encounters the target and the reference beam that impinges on the high spatial-resolution detector imparts target information---the ghost image---to the cross correlation between the photocurrents obtained from the two detectors, information that is unavailable from either photocurrent alone.  

Most ghost-imaging experiments and related theory deal with the transmissive case, in which the bucket detector is placed behind the target, and we image the target's intensity-transmission profile. In the first ghost imaging experiment, the signal and reference beams were the quantum-mechanically entangled outputs from a spontaneous parametric downconverter (SPDC) \cite{Pittman1995}, which afforded them a phase-sensitive cross correlation stronger than permitted by classical physics \cite{Erkmen2008Unified}. The system was run at low flux, in which the SPDC's post-selected output state, within the photodetectors' response time, could be taken to be a biphoton;  the ghost image was then formed by counting coincidences between the bucket and high spatial-resolution detections instead of a photocurrent cross correlation. Subsequently, ghost imaging was done with a high-flux pseudothermal source, i.e., a classical-state source realized by sending a laser beam through a rotating ground-glass diffuser and a 50-50 beam splitter to create identical, spatially-incoherent signal and reference beams with a phase-insensitive cross correlation.   The ghost image was then obtained by continuous-time cross correlation of the bucket and high spatial-resolution photocurrents \cite{Valencia2005,Ferri2005}.  Later, the theories of SPDC and pseudothermal ghost imaging were united in a Gaussian-state treatment that showed, in both cases, that ghost-image formation by photocurrent correlation arises from classical coherence propagation, but the nonclassical SPDC source offers much higher image contrast and a modest spatial-resolution advantage in near-field operation \cite{Erkmen2008Unified, Erkmen2009SNR,Erkmen2010AOP}. 

Recently, a computational ghost-imaging method was introduced, in which the need for a reference beam is removed \cite{Shapiro2008Comp,Bromberg2009}.  Conventional ghost imaging relies on the intensity cross-correlation between the signal beam on the target and the reference beam on the high spatial-resolution detector.  In SPDC and pseudothermal ghost imaging, these beam patterns are random.  If, however, a spatial light modulator is used to impart a sequence of known spatial patterns to a laser beam prior to propagation to the target, then diffraction theory can be used to calculate noiseless versions of the associated reference-beam intensity patterns.  These computed reference-beam results can be employed, for cross-correlation ghost-image formation, as if they came from photodetection of a physical reference beam.  More importantly, computational ghost imaging opens the door for more advanced computational image reconstruction techniques, notably compressive sensing \cite{Katz2009}.

All of the preceding discussion has been framed for ghost imaging of transmissive targets.  Recent experiments have shown the feasibility of an alternate ghost-imaging configuration, in which the bucket detector views the target in reflection, rather than in transmission \cite{Meyers2008}. In this reflective setup, the source and detector can, in principle, be moved arbitrarily far away from the target, suggesting that ghost imaging could be a viable remote-sensing technique.  Atmospheric turbulence will surely be a relevant concern in any such standoff-sensing application, just as it is for astronomical imaging and laser radar.  To date, there has been theoretical study of the impact of turbulence on ghost imaging in transmission \cite{Cheng2009}, but there has only been an initial turbulence-free theory development for ghost imaging in reflection \cite{Hardy2010}, and that work was limited to pseudothermal sources.  

\begin{figure}[t]
\includegraphics[width=3.5in]{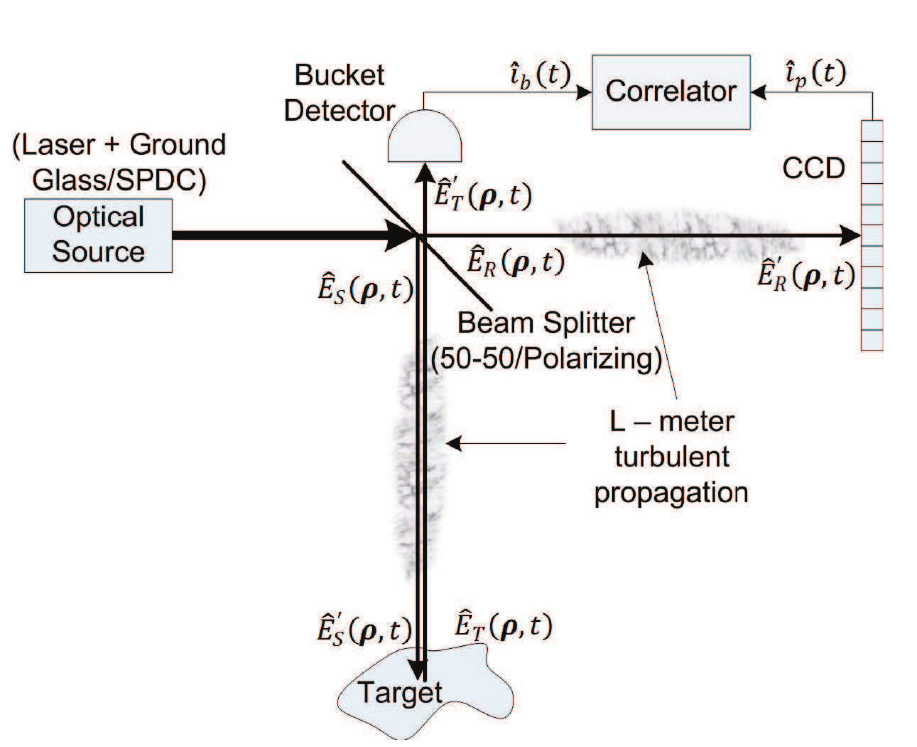}
\caption{(Color online) Light from a spatially-incoherent optical source---either a type-II phase matched SPDC or a laser rendered spatially incoherent by passage through a rotating ground-glass---is separated into signal and reference beams by a beam splitter. For the pseudothermal (laser) case, this is a 50-50 beam splitter.  For the SPDC, this is a polarizing beam splitter. The reference beam travels $L\,$m to the CCD camera, while the signal beam travels $L\,$m to the target, and the reflected light travels $L\,$m to the bucket detector. The image is then formed by cross correlation between the photocurrents obtained from the bucket detector and the high spatial-resolution (CCD camera) detector.}
\label{fig:setup}
\end{figure}

In this paper we will extend the analysis from \cite{Hardy2010} to include SPDC ghost imaging in reflection and, for both SPDC and pseudothermal operation, the presence of atmospheric turbulence in the propagation paths.  The setup we shall consider is the lensless ghost-imaging configuration shown schematically in Fig.~\ref{fig:setup} \cite{footnote1}.  Pseudothermal ghost imaging can be described quantitatively using semiclassical photodetection theory---in which the fields are described classically and photodetection incurs the shot noise associated with the discreteness of the electron charge---whereas quantum photodetection theory is required for the SPDC case, owing to the nonclassical nature of its output state \cite{Erkmen2008Unified}.  Nevertheless, to provide a unified development, we will employ quantum photodetection to characterize both pseudothermal and SPDC ghost imaging.  We begin, in Sec.~\ref{sec:Setup}, with the theoretical framework for our analysis.  Here we will describe our models for the pseudothermal and SPDC sources, reflection from the target, atmospheric propagation, quantum photodetection, and image formation. In Sec.~\ref{sec:ResCon} we will derive expressions for the spatial resolution and image contrast for pseudothermal and SPDC ghost images, and in Sec.~\ref{sec:SNR} we will derive their respective signal-to-noise ratios (SNRs). Then, in Sec.~\ref{sec:Comp}, we shall consider a computational ghost imager, in which the reference beam is calculated instead of measured \cite{Shapiro2008Comp}.  Because the reference beam is not an optical field, we find it convenient in Sec.~\ref{sec:Comp} to employ the semiclassical theory.  This section will also include a quantitative performance comparison between pseudothermal, SDPC, and computational ghost images formed in reflection.  We conclude, in Sec.~\ref{sec:Discussion}, with a discussion of our findings, and an assessment of the situations in which reflective ghost-imaging might have practical value.

\section{Setup}\label{sec:Setup}
\subsection{Optical Sources}
We are interested in ghost imaging with classical pseudothermal light, and the nonclassical output of an SPDC. In both cases our signal and reference beams will be represented quantum-mechanically as $\sqrt{\mbox{photons/m$^2$s}}$-units, $+z$-going, positive-frequency, field operators---$\E_S(\bp,t)e^{-i\omega_0t}$ and $\E_R(\bp,t)e^{-i\omega_ot}$, as functions of transverse-coordinate vector $\bp$ and time---at the output of the Fig.~\ref{fig:setup} beam splitter. Their baseband field operators have the canonical commutation relations
\begin{align}
\label{eq:Comm1}&[\E_J(\bp_1,t_1),\E_K(\bp_2,t_2)] = 0, \\[.05in]
\label{eq:Comm2}&[\E_J(\bp_1,t_1),\E^\dagger_K(\bp_2,t_2)] = \delta_{JK}\delta(\bp_1-\bp_2)\delta(t_1-t_2),
\end{align}
for $J =  S,R$ and $K = S,R$, where $\delta_{JK}$ is the Kronecker delta, and $\delta(\cdot)$ is the unit impulse. Following \cite{Erkmen2008Unified}, we will take both the pseudothermal and SPDC sources to produce signal and reference beams that are in zero-mean, jointly Gaussian states, which are then completely characterized by their nonzero autocorrelation and cross-correlation functions.  All the autocorrelations will be assumed to have the same Gaussian-Schell model form,
\begin{align}\label{eq:AutoCorr}
\avg{\E^\dagger_K&(\bp_1,t_1)\E_K(\bp_2,t_2)} \nonumber\\[.05in]
& =\frac{2P}{\pi a_0^2}e^{-\frac{|\bp_1|^2+|\bp_2|^2}{a_0^2}} e^{-\frac{|\bp_1-\bp_2|^2}{2\rho_0^2}}e^{-\frac{(t_1-t_2)^2}{2 T_0^2}},
\end{align}
for $K = S,R$, where $P$ is the photon flux, $a_0$ is the source's $e^{-2}$ intensity radius, $\rho_0$ is its coherence length, and $T_0$ is its coherence time.  The distinction between the Gaussian-state models for the pseudothermal and SPDC sources lies in their signal-reference cross correlations, as we will now explain.

The pseudothermal signal and reference beams start as continuous-wave (cw) laser light, that is passed through a rotating ground-glass diffuser to make it spatially incoherent.  Signal and reference beams are then obtained by 50-50 beam splitting. The signal and reference thus have no phase-sensitive cross correlation, and the maximum phase-insensitive cross correlation allowed by classical physics given their autocorrelation functions, i.e., 
\begin{align}
\label{eq:CCorrPS}
\avg{\E_S&(\bp_1,t_1)\E_R(\bp_2,t_2)}_C=0, \\[.05in]
\begin{split}\label{eq:CCorrPIS}
\avg{\E^\dagger_S&(\bp_1,t_1)\E_R(\bp_2,t_2)}_C \\[.05in]
& =\frac{2P}{\pi a_0^2}e^{-\frac{|\bp_1|^2+|\bp_2|^2}{a_0^2}} e^{-\frac{|\bp_1-\bp_2|^2}{2\rho_0^2}}e^{-\frac{(t_1-t_2)^2}{2 T_0^2}},
\end{split}
\end{align}
where the subscript $C$ indicates that these are classical-state cross correlations.  The coherence time is directly related to how fast the ground-glass rotates, the coherence length is typically on the order of a few wavelengths, and the intensity radius is set by the radius of the initial laser beam.

A type-II phase-matched, collinear, cw SPDC source emits orthogonally-polarized, co-propagating signal and idler beams, which are separated by a polarizing beam splitter to become the signal and reference beams for ghost imaging. They are in a maximally-entangled, jointly-Gaussian state with no phase-insensitive cross correlation, and the maximum phase-sensitive cross correlation permitted by quantum theory given their autocorrelation functions, namely \cite{Erkmen2009SNR}
\begin{widetext}
\begin{align}
\label{eq:QCorrPIS}
\avg{\E^\dagger_S(\bp_1,t_1)&\E_R(\bp_2,t_2)}_Q = 0, \\[.05in]
\begin{split}\label{eq:QCorrPS}
\avg{\E_S(\bp_1,t_1)&\E_R(\bp_2,t_2)}_Q = \frac{2P}{\pi a_0^2} e^{-\frac{|\bp_1|^2+|\bp_2|^2}{a_0^2}} \!\!
\left[e^{-\frac{|\bp_1-\bp_2|^2}{2\rho_0^2}}e^{-\frac{(t_1-t_2)^2}{2T_0^2}} + i \left(\frac{2}{\pi}\right)^\frac{1}{4} \!\sqrt{\frac{a_0^2}{P T_0 \rho_0^2}}e^{-\frac{|\bp_1-\bp_2|^2}{\rho_0^2}}e^{-\frac{(t_1-t_2)^2}{T_0^2}}  \right],
\end{split}
\end{align}
\end{widetext}
where the subscript $Q$ indicates that these are quantum (nonclassical) cross correlations.

\subsection{Propagation through Turbulence}
The signal and reference beams in Fig.~\ref{fig:setup} first propagate $L\,$m over spatially-separated paths from their shared source location \cite{footnote1}.  The reference beam is then measured by a high-spatial resolution CCD camera, while the signal field reflects off the target and travels $L\,$m to the bucket detector.  We will assume statistically-independent turbulence on all three paths:  the reference path ($R$) from the source to the CCD; the signal path ($S$) from the source to the target; and the target-return path ($T$) from the target to the bucket detector. 

To account for the Kolmogorov-spectrum turbulence that is uniformly distributed along the three propagation paths, we make use of the quantum version of the extended Huygens-Fresnel principle \cite{Strohbehn1978,footnote2}
\begin{align}\label{eq:Prop}
\E'_m(\bp',t) = \int\! d\bp\, \E_m(\bp,t)  \frac{k_0 e^{ik_0(L+|\bp'-\bp|^2/2L)}}{i2\pi L}\,e^{\psi_m(\bp', \bp)},
\end{align}
where $k_0 = \omega_0/c = 2\pi/\lambda_0$, and $\psi_m(\bp', \bp)$ is a complex-valued random process that encapsulates the effects of turbulence on path $m$ from $\bp$ to $\bp'$.  The real and imaginary parts of $\psi_m(\bp',\bp)$, which we will denote $\chi_m(\bp',\bp)$ and $\phi_m(\bp',\bp)$, are the logamplitude and phase fluctuations imposed on the field received at transverse coordinate $\bp'$ over path $m$ from a point source at transverse coordinate $\bp$.  Because all the path lengths in Fig.~\ref{fig:setup} are equal, we have suppressed the $L/c$ line-of-sight propagation delay in Eq.~(\ref{eq:Prop}).  We have also suppressed any time-dependence of the turbulence.  Physically, this amounts to assuming that the integration time, $T_I$, employed in the Fig.~\ref{fig:setup} correlator is shorter than the $\sim$ms coherence time of the turbulence.  However, that assumption is \em not\/\rm\ necessary for deriving the spatial resolution and image contrast achieved in the presence of turbulence, although it \em will\/\rm\ be relevant to our signal-to-noise ratio analysis.  

The mutual coherence function of $e^{\psi_m(\bp',\bp)}$ is taken to be  \cite{Ishimaru1978,Osche2002}
\begin{align}\label{eq:TurbCorr}
\avg{&e^{\psi^*_m(\bp'_1, \bp_1)}e^{\psi_m(\bp'_2, \bp_2)}}\nonumber \\[.05in] &=e^{-(|\bp'_1-\bp'_2|^2+(\bp'_1-\bp'_2)\cdot(\bp_1-\bp_2) +|\bp_1-\bp_2|^2)/2\rho_m^2},
\end{align}
which is the square-law approximation to the rigorous $5/3$-law behavior for uniformly-distributed turbulence \cite{footnote3}.  In Eq.~(\ref{eq:TurbCorr}), $\rho_m$ is the turbulence coherence length on path $m$, given by
\begin{align}\label{eq:TurbCL}
\rho_m = (1.09 k_0^2 C_{n,m}^2 L)^{-3/5},
\end{align}
in terms of the constant turbulence strength-parameter, $C^2_{n,m}$, along path $m$ \cite{Ishimaru1978}.

We  will assume that the path length $L$ satisfies the far-field coherence-propagation conditions for pseudothermal and SPDC light discussed in \cite{Erkmen2008Unified,Erkmen2010AOP}, viz.,  $k_0 a_0 \rho_0/2L \ll 1$, and $k_0 a_0^2/2L \ll 1$, respectively.  For convenience, we also define $\rho_L = 2L/k_0 a_0$ and $a_L = 2L/k_0\rho_0$.  These are the far-field, turbulence-free, on-target coherence length and intensity radius produced by the Gaussian-Schell autocorrelation function from Eq.~(\ref{eq:AutoCorr}).

\subsection{Target Reflection}
Most real-world targets have surfaces that are sufficiently rough---on the scale of an optical wavelength---that reflections from them present fully-developed speckle patterns at appreciable distances from the surface \cite{Goodman2007}. We will model such a target as a planar 2-D object with random, microscopic height variations---having depths on the order of several wavelengths, and transverse variations on the order a wavelength---and an average intensity-reflection coefficient $\sT(\bp)$.  Thus, following laser-radar theory \cite{Shapiro1981}, we shall assume that this target has a random field-reflection coefficient, $T(\bp)$, that is a zero-mean, complex-valued Gaussian random process characterized by the autocorrelation function
\begin{align}\label{eq:Scatter}
\avg{T^*(\bp_1)T(\bp_2)} = \lambda_0^2 \sT(\bp_1) \delta(\bp_1-\bp_2),
\end{align}
where the deterministic pattern $\sT(\bp)$ is what we are tying to image. 

Target reflection is a passive process, so we require $|T(\bp)| \le 1$, which is in conflict with the Gaussian statistics and the delta-function term in Eq.~(\ref{eq:Scatter}).  That delta function, however, leads to quasi-Lambertian reflection, implying that at standoff distances a realistic bucket detector will only capture a very small fraction of the reflected light.  Hence using the reflection statistics from the preceding paragraph is a reasonable approximation for ghost imaging of rough-surfaced targets.  However, because some light may be absorbed by or transmitted through the target, the quantum model for target reflection needs to be 
\begin{align}\label{eq:TargetLoss}
\E_T(\bp,t) = T(\bp)\E'_S(\bp,t) +\sqrt{1-|T(\bp)|^2}\,\E_{vac,S}(\bp,t)
\end{align}
Here $\E_{vac,S}(\bp,t)$ is a vacuum-state field operator, needed to preserve the free-field commutator relations given, for the source fields, in Eqs.~\eqref{eq:Comm1} and \eqref{eq:Comm2}.  

In what follows we will assume that the target's intensity-reflection coefficient, $\sT(\bp)$, is space-limited to the region illuminated by the signal beam, i.e., $|\sT(\bp)| = 0$ for $|\bp| > a_L$.  This condition will give the ensemble-average pseudothermal and SDPC ghost images convolution-integral relationships with $\sT(\bp)$ from which to determine their spatial resolutions.  See \cite{Erkmen2008Unified,Erkmen2010AOP} for information about ghost imaging field-of-view when the preceding restriction on $\sT(\bp)$ is not satisfied.

\subsection{Photodetection and Image Formation}
Loss also occurs in the photodetection process, because the detectors in Fig.~\ref{fig:setup} will, in general, have sub-unity quantum efficiencies.  Therefore, the field operator $\E_p(\bp,t)$ driving the photocurrent, $\hat{i}_p(t)$, from  CCD pixel $p$, and the field operator $\E_b(\bp,t)$ driving the photocurrent, $\hat{i}_b(t)$, from the bucket detector satisfy \cite{footnote4,Shapiro2009}
\begin{align}\label{eq:Detector}
\E_m(\bp,t) = \sqrt{\eta}\,\E'_\ell(\bp,t) +\sqrt{1-\eta}\,\E_{vac,\ell}(\bp,t),
\end{align}
for $(m,\ell)=(p,R)$ or $(m,\ell)=(b,T)$. Here, $\eta$ is the quantum efficiency, which is taken to be the same for both the bucket detector and the CCD camera, and $\E_{vac,\ell}(\bp,t)$ is a vacuum-state field operator whose inclusion models the loss incurred by having $\eta < 1$, cf.\ Eq.~(\ref{eq:TargetLoss}).

We will assume ghost-image formation via continuous-time, pixel-wise correlation of the photocurrents from the CCD and the bucket detector, formalized as
\begin{align}\label{eq:Corr}
\hat{C}(\bp_p) = \frac{1}{T_I}\int_{-T_I/2}^{T_I/2}\!dt\, \hat{i}_p(t) \hat{i}_b(t),
\end{align}
where $\bp_p$ is the center coordinate of the CCD's $p$th pixel, and $T_I$ is the correlation integration time.  From quantum photodetection theory \cite{Shapiro2009}, we have that
\begin{align}\label{eq:Current}
\hat{i}_m(t) = q\int d\tau\, \hat{P}_m(\tau) h(t-\tau), \quad m=p,b
\end{align}
where 
\begin{align}\label{eq:Flux}
\hat{P}_m(t) = \int_{{\cal A}_m} d\bp\, \E^\dagger_m(\bp,t) \E_m(\bp,t)
\end{align}
is the effective (sub-unity quantum efficiency) photon flux on the active region ${\cal A}_m$ for detector $m$, and $h(t)$ is the impulse response of the detectors' output circuit, which we have taken to be the same for both the bucket detector and the CCD camera.

To enable the performance evaluations in the sections to follow, we make the following assumptions. First, each pixel is small enough that the the average photon flux impinging on it obeys $\langle \hat{P}_p(t)\rangle  = A_p\langle \E^\dagger_p(\bp_p,t)\E_p(\bp_p,t)\rangle$, where $A_p$ is the area of the ${\cal A}_p$.   Second, we assume that the detector has a Gaussian impulse response, 
\begin{align}
h(t) = \frac{\Omega_B}{\sqrt{8\pi}}e^{-\Omega^2_B t^2/8},
\end{align}
whose bandwidth, $\Omega_B$, greatly exceeds that of the incident light for the pseudothermal source ($\Omega_B T_0 \gg 1$), and is much smaller than that of the incident light for the SPDC source ($\Omega_B T_0 \ll 1$). Finally, we assume that the correlation integration time is sufficiently long to capture a ghost image, so $T_I \gg T_0$ for the pseudothermal source, and $T_I \gg \Omega_B^{-1}$ for the SPDC source. 
 
\section{Resolution and Contrast}\label{sec:ResCon}
The spatial resolution and contrast of the ghost image are properties of the average cross correlation, $\langle C(\bp_p)\rangle$.  To determine this ensemble-average cross correlation we: (1)  write $\hat{i}_b(t)$ and $\hat{i}_p(t)$ in terms of the field operators that drive them, $\E_b(\bp,t)$ and $\E_p(\bp,t)$; (2) apply the canonical commutator relations to normally-order the resulting photocurrent-product expressions; and (3) use Eq.~\eqref{eq:Detector} to write the detected fields in terms of $\E'_R(\bp,t)$ and $\E'_T(\bp,t)$. From the linearity of expectation, we can bring the ensemble-average operation inside the time-domain integration to get
\begin{align}\label{eq:AvgCorr}
\avg{\hat{C}(&\bp_p)} = q^2\eta^2 A_p \int\!d\tau_1\int\!d\tau_2 \int_{{\cal A}_b} \! d\bp \, h(t-\tau_1)h(t-\tau_2) \nonumber \\[.05in]
&\times \avg{\E'^\dagger_R(\bp_p,\tau_1) \E'^\dagger_T(\bp,\tau_2) \E'_R(\bp_p,\tau_1) \E'_T(\bp,\tau_2)}.
\end{align}
We then use Eqs.~\eqref{eq:Prop} and \eqref{eq:TargetLoss} to back-propagate the target-return field operator to the target, and find
\begin{eqnarray}
\lefteqn{
\avg{\E'^\dagger_R(\bp_p,\tau_1) \E'^\dagger_T(\bp,\tau_2) \E'_R(\bp_p,\tau_1) \E'_T(\bp,\tau_2)}}
\nonumber  \\[.05in]
&=&\int\! d\bp_2 \!\int\! d\bp_3\, \avg{T^*(\bp_2)T(\bp_3)}\avg{e^{\psi^*_T(\bp,\bp_2)}e^{\psi_T(\bp,\bp_3)}}\nonumber \\[.05in]
&\times& \avg{\E'^\dagger_R(\bp_p,\tau_1) \E'^\dagger_S(\bp_2,\tau_2) \E'_R(\bp_p,\tau_1) \E'_S(\bp_3,\tau_2)} \nonumber \\[.05in]
&\times& \frac{e^{-ik_0(|\bp_2|^2-|\bp_3|^2)/2L + ik_0\bp\cdot(\bp_2-\bp_3)/L}}{(\lambda_0L)^2},
\end{eqnarray}
where we have used the statistical independence of the target's reflection coefficient, the signal and reference fields at the source, and the turbulence on all three propagation paths.

Substituting for $ \avg{T^*(\bp_2)T(\bp_3)}$ from Eq.~\eqref{eq:Scatter} and performing the $\bp_2$ integral eliminates the turbulence term from the preceding $\langle C(\bp_p)\rangle$ expression, which shows that turbulence on the the target-to-bucket path has no impact on the ghost image's spatial resolution or image contrast.  Now, by back-propagating the field operators to the source---by means of Eq.~\eqref{eq:Prop} with $k_0a_0^2/2L \ll 1$---we have that
\begin{eqnarray}
\lefteqn{\avg{\E'^\dagger_R(\bp_p,\tau_1) \E'^\dagger_T(\bp,\tau_2) \E'_R(\bp_p,\tau_1) \E'_T(\bp,\tau_2)} }\nonumber \\[.05in]
&=&\frac{1}{\lambda_0^4L^6}\int\! d\bp_2 \, \sT(\bp_2) \int\! d\bp'_1 \int\!d\bp'_2 \int\!d\bp''_1 \int \! d\bp''_2 \nonumber \\[.05in]
&\times& \avg{\E^\dagger_R(\bp'_1,\tau_1) \E^\dagger_S(\bp'_2,\tau_2) \E_R(\bp''_1,\tau_1) \E_S(\bp''_2,\tau_2)} \nonumber \\[.05in]
&\times& \avg{e^{\psi^*_R(\bp_p,\bp'_1)}e^{\psi_R(\bp_p,\bp''_1)}}\avg{e^{\psi^*_S(\bp_2,\bp'_2)}e^{\psi_S(\bp_2,\bp''_2)}} \nonumber \\[.05in]
&\times& e^{ik_0[\bp_p\cdot(\bp'_1-\bp''_1) + \bp_2\cdot(\bp'_2 - \bp''_2)]/L},
\end{eqnarray}
where we have again invoked the statistical independence of the signal and reference fields at the source and the turbulence on the source-to-CCD, and source-to-target paths.

We can now directly evaluate all remaining moments. The turbulence moments are available from Eq.~\eqref{eq:TurbCorr}, and the fourth-order field moment can be expressed in terms of second-order moments with the Gaussian-state moment-factoring theorem \cite{Shapiro1994}: 
\begin{eqnarray}
\lefteqn{\hspace*{-.1in}\avg{\E^\dagger_R(\bp'_1,\tau_1) \E^\dagger_S(\bp'_2,\tau_2) \E_R(\bp''_1,\tau_1) \E_S(\bp''_2,\tau_2)} =  } \nonumber \\
&& \hspace{-.1in}\avg{\E^\dagger_R(\bp'_1,\tau_1) \E^\dagger_S(\bp'_2,\tau_2)} \avg{\E_R(\bp''_1,\tau_1) \E_S(\bp''_2,\tau_2)}+ \nonumber \\
&& \hspace*{-.1in}\avg{\E^\dagger_R(\bp'_1,\tau_1)\E_R(\bp''_1,\tau_1)} \avg{\E^\dagger_S(\bp'_2,\tau_2)\E_S(\bp''_2,\tau_2)} + \nonumber \\
&& \hspace*{-.1in} \avg{\E^\dagger_R(\bp'_1,\tau_1)\E_S(\bp''_2,\tau_2)} \avg{\E^\dagger_S(\bp'_2,\tau_2)\E_R(\bp''_1,\tau_1)}.
\label{eq:MomentFactor}
\end{eqnarray}

Until this point, our analysis applies equally well to the pseudothermal and SPDC ghost imagers. That equivalence disappears, however, when we evaluate the second-order moments in Eq.~\eqref{eq:MomentFactor} using Eqs.~\eqref{eq:AutoCorr}--\eqref{eq:CCorrPIS} for the pseudothermal ghost imager, and using Eqs.~\eqref{eq:AutoCorr}, \eqref{eq:QCorrPIS}, and \eqref{eq:QCorrPS} for the SPDC ghost imager. The results we obtain are
\begin{eqnarray}
\lefteqn{\avg{\hat{C}(\bp_p)}_C = \frac{q^2\eta^2 A_p A_b}{L^2}\left(\frac{2P}{\pi a_L^2}\right)^2} \nonumber \\
&\times& \int\!d\bp\, \sT(\bp)\left[1+\frac{ e^{-|\bp-\bp_p|^2/\alpha\rho_L^2}}{\alpha}\right],
\label{eq:CAvgMean}
\end{eqnarray}
for the pseudothermal (classical) imager, and
\begin{eqnarray}
\lefteqn{\avg{\hat{C}(\bp_p)}_Q= \frac{q^2\eta^2A_p A_b}{L^2} \left(\frac{2P}{\pi a_L^2}\right)^2\int \! d\bp\, \sT(-\bp)} \nonumber \\[.05in]
&\times& \left[ 1+ \frac{\Omega_B T_0}{4}\frac{e^{-|\bp-\bp_p|^2/\alpha\rho_L^2}}{\alpha}\left(1+\frac{1}{4\sqrt{\pi}\mathcal{I}}\right)\right].
\label{eq:QAvgMean}
\end{eqnarray}
for the SPDC (quantum) ghost imager, where $A_b$ is the area of $\mathcal{A}_b$, $\mathcal{I}\equiv P T_0 \rho_0^2/a_0^2 = PT_0\rho_L^2/a_L^2$ is the source brightness in photons per spatiotemporal mode, and
\begin{align}
\alpha \equiv \frac{2\rho_R^2\rho_S^2+a_0^2(\rho_R^2 +\rho_S^2)}{2\rho_R^2\rho_S^2}
\end{align}
is a turbulence-induced performance-degradation factor. We see that both average correlation functions are the sum of the same featureless background term
\begin{align}\label{eq:C0}
C_0=\frac{q^2\eta^2 A_p A_b}{L^2}\left(\frac{2P}{\pi a_L^2}\right)^2 \int\!d\bp\, \sT(\bp),
\end{align}
and different image-bearing terms:
\begin{align}
C_{1,C}&(\bp_p)=\frac{q^2\eta^2 A_p A_b}{\alpha L^2}\left(\frac{2P}{\pi a_L^2}\right)^2 \int\!d\bp\, \sT(\bp) e^{-\frac{|\bp-\bp_p|^2}{\alpha\rho_L^2}},
\end{align}
for the classical-state (pseudothermal) source, and
\begin{eqnarray}
C_{1,Q}(\bp_p)&=&\frac{q^2\eta^2 A_p A_b}{\alpha L^2}\left(\frac{2P}{\pi a_L^2}\right)^2\! \left(1+\frac{1}{4\sqrt{\pi}\mathcal{I}}\right) \nonumber\\[.05in]
&\times& \frac{\Omega_B T_0}{4}\int\!d\bp\, \sT(-\bp) e^{-\frac{|\bp-\bp_p|^2}{\alpha\rho_L^2}},
\end{eqnarray}
for the nonclassical-state (quantum) source.

The average pseudothermal ghost image is erect, and consists of a scaled version of the target's intensity-reflection coefficient $\sT(\bp)$ convolved with a Gaussian point-spread function whose $e^{-1}$ radius is $\rho_L\sqrt{\alpha}$. The average SPDC ghost image is inverted, with a different scaling but the same point-spread function.  Thus, both the pseudothermal and SPDC ghost images have the same spatial resolution, 
\begin{eqnarray}
\lefteqn{\rho_L\sqrt{\alpha} =  \frac{\lambda_0 L}{\pi a_0}\sqrt{1+ \frac{a_0^2}{2}\left[\rho_S^{-2} + \rho_R^{-2}\right]} = \label{eq:Res}}
\\[.05in]
&&\hspace*{-.1in}
 \frac{\lambda_0L} {\pi a_0}\sqrt{1+\frac{a_0^2}{2} (1.09 k_0^2 L)^\frac{6}{5}\left[(C_{n,S}^2)^\frac{6}{5} + (C_{n,R}^2)^\frac{6}{5}\right]},
\end{eqnarray}
where we have utilized Eq.~\eqref{eq:TurbCL} to write the turbulence coherence lengths on each path in terms of their respective strength parameters. 
Equation~\eqref{eq:Res} shows that turbulence-limited resolution prevails for the far-field lensless ghost image when either the source-to-target coherence length ($\rho_S$) or the source-to-reference coherence length ($\rho_R$) becomes smaller than the source size ($a_0$).  When $\rho_S \gg a_0$ and $\rho_R \gg a_0$, far-field lensless ghost imaging in reflection yields the same $\lambda_0L/\pi a_0$ spatial resolution previously reported for far-field lensless ghost imaging in transmission \cite{Erkmen2008Unified, Hardy2010}.  However, even when turbulence does not degrade ghost-image spatial resolution, there is still a significant difference between the transmissive and reflective cases, namely the $A_b/L^2$ factor that is present in our results but is absent from those in \cite{Erkmen2008Unified}.  This factor, which will obey $A_b/L^2 \ll 1$ in a standoff-sensing scenario, is the bucket detector's angular subtense at the target, and represents the fraction of the quasi-Lambertian reflected light that this detector collects.  No such factor appeared in \cite{Erkmen2008Unified}, because that work assumed the bucket detector collected all the light that was transmitted through the target.   

The featureless background terms in our pseudothermal and SPDC ghost images affect the images' contrast.  To assess these contrasts, we will adopt the definition and approach presented in \cite{Erkmen2008Unified}.   In particular, the image contrast $\mathcal{C}$ is the ratio of the difference between the brightest and darkest pixels in the image to the featureless background:
\begin{align}\label{eq:ContrastDef}
\mathcal{C} = \frac{\max_{\mathcal{R}}[\hat{C}(\bp_p)]-\min_{\mathcal{R}}[\hat{C}(\bp_p)]}{C_0},
\end{align}
where $\mathcal{R}$ is the region within which $\sT(\bp)$ is nonzero.
We will assume that the ghost imager completely resolves the target, so that we can use the approximation
\begin{align}\label{eq:Resolve}
\int\!d\bp\,\sT(\pm\bp)e^{-\frac{|\bp-\bp_p|^2}{\alpha\rho_L^2}} \approx \pi \alpha \rho_L^2 \sT(\pm\bp) .
\end{align}
Defining $A_T=\int\!d\bp\, \sT(\bp)$ as the target's reflective cross-section, and taking $\max_{\mathcal{R}}[\sT(\bp)] = 1$, we get the following image contrast expressions for our two ghost imagers, 
\begin{align}
\label{eq:CContrast}
&\mathcal{C}_C = \frac{\pi \rho_L^2}{A_T},
\end{align}
and
\begin{align}
\label{eq:QContrast}
&\mathcal{C}_Q = \frac{\pi \rho_L^2}{A_T}\frac{\Omega_BT_0}{4}\left(1+\frac{1}{4\sqrt{\pi}\mathcal{I}} \right).
\end{align}  

If $|\sT(\bp)| \sim 1$ where it is nonzero, then $A_T$ is the area of the target.  Moreover, $\rho_L^2$ is the spatial-resolution area in the absence of turbulence.  It follows that the pseudothermal ghost-image contrast is approximately $1/N$, where $N$ is number of no-turbulence resolution cells in the image. Note that our analysis to this point has assumed a {\sc dc}-coupled correlator in Fig.~\ref{fig:setup}.  The contrast of a pseudothermal ghost image can be improved by suppressing the background term through {\sc ac}-coupling, i.e., measuring the cross covariance, rather than the cross correlation, between the bucket and high spatial-resolution photodetector's outputs, as was done in the pseudothermal experiments of Scarcelli \em et al\/\rm.\ \cite{Scarcelli2006}.  

The pseudothermal imager's contrast does not depend on its source brightness, $\mathcal{I}$, but that for the SPDC does.   At high source brightness, $\mathcal{I} \gg 1$, the SPDC image contrast reduces to the result we found for the pseudothermal case.  The normal operating regime for a cw SPDC, however, is one of low brightness, $\mathcal{I} \ll 1$, for which Eq.~\eqref{eq:QContrast} reduces to 
\begin{equation}
\mathcal{C}_Q \approx \frac{\sqrt{\pi}}{16}\,\frac{\Omega_B}{P}\frac{a_L^2}{A_T}.
\end{equation}
The number of photon pairs emitted by an SPDC in one detector time-constant is low, so $\Omega_B /P \gg 1$, and our assumption that the target is contained within the region illuminated by the signal beam implies that $a_L^2/A_T > 1$ if $|\sT(\bp)| \sim 1$ where it is nonzero.  Consequently, unlike the pseudothermal case, $\mathcal{C}_Q \gg 1$ is the norm for SDPC ghost imaging in {\sc dc}-coupled operation.

\section{Signal-to-Noise Ratio}\label{sec:SNR}
In keeping with the analysis done in \cite{Erkmen2009SNR}, we will be evaluating the signal-to-noise ratio (SNR) pixel-wise as the ratio of the square of the expected value to the variance for the photocurrent cross-correlation function,
\begin{align}\label{eq:SNRi}
\text{SNR} = \frac{\avg{\hat{C}(\bp_p)}^2}{{\rm Var}[\hat{C}(\bp_p)]} = \frac{\avg{\hat{C}(\bp_p)}^2}{\avg{\hat{C}^2(\bp_p)}-\avg{ \hat{C}(\bp_p)}^2}.
\end{align}
The main complication in evaluating Eq.~\eqref{eq:SNRi} is the second moment,
\begin{align}\label{eq:Cm2}
\avg{\hat{C}^2(\bp_p)} = \frac{1}{T_I^2}\!\int\limits_{-T_I/2}^{T_I/2}\!\!\!dt\!\!\int\limits_{-T_I/2}^{T_I/2}\!\!\!du\, \avg{\hat{i}_p(t)\hat{i}_b(t)\hat{i}_p(u)\hat{i}_b(u)}.
\end{align}
After using Eq.~\eqref{eq:Current} to write the currents in terms of the fields, Eqs.~\eqref{eq:Comm1} and \eqref{eq:Comm2} to normally order the fields, and Eqs.~\eqref{eq:Prop} and \eqref{eq:TargetLoss} to propagate the fields back to the source, we are left needing to evaluate an eighth-order moment of the fields, two sixth-order moments of the fields, a fourth-order moment of the fields, a fourth-order moment of the target, three fourth-order moments of the  turbulence, and propagation of these terms over 12 unique paths. To make the analysis more tractable, we use several simplifying assumptions. First, we {\sc ac}-couple both of the detector outputs by inserting a narrowband notch centered at the origin in frequency space, resulting in the frequency response
\begin{align}\label{eq:DCblock}
H_B(\Omega) = e^{-2\Omega^2/\Omega^2_B}-e^{-2\Omega^2/\Omega^2_N},
\end{align}
where $\Omega_N \ll \Omega_B$.
We also assume that the ghost imager can resolve all structure on the target, allowing us to use the approximation in Eq.~\eqref{eq:Resolve}. 

To evaluate the fourth-order turbulence moments, we can no longer use the square-law approximation in Eq.~\eqref{eq:TurbCorr} that models the turbulence's mutual-coherence function as Gaussian, because this approximation is not valid for fourth-order moments \cite{Wandzura1980}. To evaluate these terms we assume that the turbulence coherence length on each path is much larger than the on-target coherence length $\rho_L$. We also assume the turbulence coherence lengths are all much larger than $a_0$, implying that the turbulence functions are independent of coordinates at the source and detection planes. Although these assumptions are somewhat limiting, they still allow weak-to-medium strength turbulence for a typical situation.  For instance, if the target is standing off $1\,$km, the source aperture has radius $a_0=3\,$cm, and we are at a wavelength $\lambda_0=1.5\,\mu$m, this assumption is satisfied for $C_n^2 \le 10^{-14}\,{\rm m}^{-2/3}$. Moreover, this situation has no spatial-averaging of the turbulence at the target, and thus represents a worst-case scenario for the SNR.

For $\bp'_\ell$ and $\bp''_\ell$ being coordinates at the CCD or bucket detectors, and $\bp_\ell$ being coordinates at the target, the preceding coherence-length assumptions imply that
\begin{align}
\avg{e^{\psi^*_m(\bp'_1, \bp_1)}&e^{\psi_m(\bp''_1, \bp_1)}e^{\psi^*_m(\bp'_2, \bp_2)}e^{\psi_m(\bp''_2, \bp_2)}}\nonumber \\
 &\approx \avg{e^{\psi^*_m(\bm 0, \bp_1)}e^{\psi_m(\bm 0, \bp_1)}e^{\psi^*_m(\bm 0, \bp_2)}e^{\psi_m(\bm 0, \bp_2)}}\nonumber \\
  &= \avg{e^{2{\rm Re}[\psi_m(\bm 0, \bp_1)]+2{\rm Re}[\psi_m(\bm 0, \bp_2)]}}.
\end{align}
Decomposing the turbulence function into its log-amplitude and phase components as $\psi_m(\bp_1,\bp_2)=\chi_m(\bp_1,\bp_2)+i\phi_m(\bp_1,\bp_2)$, and taking the logamplitude to be Gaussian distributed with mean equal to minus its variance \cite{Strohbehn1978}, then gives 
\begin{eqnarray}
\avg{e^{\psi^*_m(\bp'_1, \bp_1)}e^{\psi_m(\bp''_1, \bp_1)}e^{\psi^*_m(\bp'_2, \bp_2)}e^{\psi_m(\bp''_2, \bp_2)}} \nonumber \\[.05in]
\approx e^{4 K_{\chi_m} (\bp_1-\bp_2)},
\end{eqnarray}
where 
\begin{equation}
K_{\chi_m} (\bp_1-\bp_2) = \avg{\Delta\chi_m (\bm 0, \bp_1)\Delta\chi_m (\bm 0, \bp_2)},
\end{equation}
for $\Delta\chi_m \equiv \chi_m - \avg{\chi_m}$, is the log-amplitude covariance function.  The logamplitude variance will be taken to be \cite{Osche2002}
\begin{align}
\sigma_m^2=K_{\chi_m} (\bm 0)=0.124 \,C_{n,m}^2 k_0^{7/6}L^{11/6},
\end{align}
where we have used the Rytov-approximation expression, whose validity is ensured by our assumption of weak-to-medium strength turbulence.

Even with these simplifications, the SNR evaluation is quite tedious, if somewhat straightforward. The higher-order field moments and the fourth-order target moment can be reduced to a sum of the products of second-order moments by application of the Gaussian moment-factoring theorem. The {\sc ac}-coupling removes several of these terms, and we are left with integrals and Fourier transforms of the product of Gaussian terms, the turbulence correlation functions, and the target's intensity-reflection coefficient, $\sT(\bp)$. To simplify our final SNR expressions, we define two new terms,
\begin{align}
\label{eq:ATp}A'_T & = \int\! d\bp\, \sT^2(\bp), \\
\label{eq:Gamma}\Gamma &= \frac{1}{(4\pi\beta)^2}\int \! d\bm\nu\, \ e^{-|\bm\nu|^2/2}O(\bm\nu,4\sqrt{\beta}),
\end{align}
where $O(\bm\zeta, D)$ is the two-circle overlap function for circles of diameter $D$, 
\begin{eqnarray}
\lefteqn{\hspace*{-.1in}O(\bm\zeta,D) = } \nonumber \\[.05in]
&&\hspace*{-.2in}\left\{\begin{array}{ll}
\frac{D^2}{2}\left[ \cos^{-1} \left( \frac{|\bm\zeta|}{D}\right) -\frac{|\bm\zeta|}{D}\sqrt{1-\frac{|\bm\zeta|^2}{D^2}} \right], & \mbox{for $|\bm\zeta|\le D$} \\[.05in]
0, & \text{elsewhere,}
\end{array}\right.
\end{eqnarray}
$\beta = A_b/\pi a_0^2$, and we have utilized the dimensionless difference coordinates $\bm{\nu} = \rho_Lk_0(\bm{\rho}'-\bm{\rho}'')/L$, where $\bp'$ and $\bp''$ are coordinates at the bucket detector. 

Both $A'_T$ and $\Gamma$ have significant physical interpretations. Similar to $A_T$, $A'_T$ is another measure of the reflective area of the target, and thus directly related to the number of on-target resolution cells, and the subsequent amount of time it takes to form an image. $\Gamma$ is a metric for the spatial averaging of the target-induced speckle over the bucket detector. For very small detectors $\Gamma \approx 1$, while for large detectors it is inversely proportional to the area of the bucket detector, viz.  $\Gamma \approx 1/2\beta$. To further simplify our SNR expressions, we make two final assumptions: first that $A'_T/\rho_L^2\gg 30$, which is equivalent to saying that the ghost image consists of at least $10 \times 10$ resolution cells; and second, that the bucket detector is at least as large as the area of the source beam, or $\beta \ge 1$. Applying these conditions we arrive at the following SNR expressions for pseudothermal and SPDC reflective ghost imaging through atmospheric turbulence,
\begin{widetext}
\begin{gather}
\label{eq:CSNR}
\text{SNR}_C = \frac{\displaystyle \sT^2(\bp_p)}{\displaystyle\frac{T_0}{T_I}\Delta^2\text{Source} + \Delta^2\text{Path} + \frac{\Omega_BT_0^2}{T_I}\Delta^2\text{Detect}+\frac{T_0}{T_I}\Delta^2\text{Mix}} \\
\label{eq:QSNR}
\text{SNR}_Q = \frac{\displaystyle \sT^2(\bp_p) \left[1+\frac{1}{4\sqrt{\pi}\mathcal{I}}\right]^2}{\displaystyle \frac{4}{\Omega_BT_I}\Delta^2\text{Source} + \Delta^2\text{Path}\left[1+\frac{1}{4\sqrt{\pi}\mathcal{I}}\right]^2 + \frac{T_0 4\sqrt{2}}{T_I}\Delta^2\text{Detect}\left[1+\frac{1}{4\sqrt{\pi}\mathcal{I}}\right]+ \frac{T_0}{T_I}\Delta^2\text{Mix}\left[\frac{2}{\sqrt{3}}+\frac{1}{4\sqrt{\pi}\mathcal{I}}\right]}.
\end{gather}
\end{widetext}
where, for ease of comparison with the pseudothermal imager, we have inverted the image coordinates in the SPDC case.  

The terms that appear in the noise denominators of Eqs.~(\ref{eq:CSNR}) and (\ref{eq:QSNR}) are as follows:  
\begin{align}
&\Delta^2\text{Source} = \frac{A'_T (1+\beta^{-1})e^{4(\sigma_R^2+\sigma_S^2+\sigma_T^2)}}{\sqrt{2\pi}\rho_L^2} , \\
&\Delta^2\text{Path} = \sT^2(\bp_p)\left[e^{4(\sigma_R^2+\sigma_S^2+\sigma_T^2)}\left(\Gamma+1\right)-1\right], \\
&\Delta^2\text{Detect} =\frac{\sT(\bp_p) \rho_L^2 \sqrt{\pi}}{16 \sqrt{2} A_p \eta^2 \mathcal{I}^2}\frac{L^2}{A_b}, \\
&\Delta^2\text{Mix} = \frac{\sT(\bp_p)}{\eta\mathcal{I}}\frac{ L^2}{A_b}e^{4\sigma_R^2}\nonumber\\
&\quad\quad+\frac{\pi\rho_L^2\sT^2(\bp_p)}{A_p\eta\mathcal{I}}\left[\frac{4}{3}+\beta^{-1}\right]e^{4(\sigma_S^2+\sigma_T^2)},
\end{align}
where $\sigma^2_\ell$ for $\ell = R, S, T$ denotes the turbulence-induced logamplitude variance on the reference ($R$), signal ($S$), and target ($T$) paths.  Each of these noise terms has an important physical interpretation, as we explain below. 

The first noise term, $\Delta^2\text{Source}$, captures the fluctuations associated with decorrelation of transverse locations in the image arising from the source's time evolution. Thus, it is inversely proportional to the number of on-target resolution cells, $\rho_L^2/A'_T$, and the number of different source patterns measured. For the pseudothermal imager, for which $\Omega_BT_0 \gg 1$ holds, this is the number of field-coherence times in the integration time, so we find $\Delta^2\text{Source}$ scaled by $T_0/T_I$. For the SPDC imager we have $\Omega_BT_0 \ll 1$, so it is the detectors' response time that limits the measurements' correlation behavior, hence we find $\Delta^2\text{Source}$ scaled by $4/\Omega_BT_I$ in that imager's SNR denominator.

The $\Delta^2\text{Path}$ term is due to the optical-path interactions that degrade the measurement, i.e., the turbulent atmosphere, and the scattering off the rough target. The noise contribution from target scattering can be mitigated by aperture averaging, which, as noted above, is quantified by $\Gamma$. The effect of the turbulence is somewhat more complicated, as it contributes a scaling factor $e^{4\sigma_\ell^2}$ to all field-variation terms on path $\ell$. In addition, the logamplitude fluctuations also cause a deviation that results in the squared-mean term in the denominator of Eq.~\eqref{eq:SNRi} no longer being canceled. These effects are particularly devastating because $\Delta^2\text{Path}$ is the only noise term whose strength is independent of the detector integration time.   This is because we have assumed time-independent target behavior and $T_I$ less than the $\sim$ms turbulence coherence time.  Consequently, regardless of the source brightness and integration time, $\Delta^2\text{Path}$ sets an ultimate upper limit on the SNR.

The noise term $\Delta^2\text{Detect}$ results from the statistics of photodetection at both the reference and bucket detectors, and as such is inversely proportional to the photon fluxes impinging on their active regions.   Thus, $\Delta^2\text{Detect}$ is inversely proportional to the size of each pixel, $A_p$, and the solid-angle subtense factor $A_b/L^2$. Detection noise is mitigated by temporal averaging, but exacerbated by having a fast detector, so the pseudothermal $\Delta^2\text{Detect}$ is scaled by $\Omega_BT_0^2/T_I$, while for the SPDC imager it is scaled only by $T_0/T_I$. The SPDC case is also scaled by by $1+1/4\sqrt{\pi}\mathcal{I}$, a result of that source's stronger-than classical correlation. 

The $\Delta^2\text{Mix}$ term contains noise contributed by mixing between source plus path noise on the reference arm with detection noise on the signal arm, and vice versa. For both imagers this term has a temporal scaling of $T_0/T_I$, and for the SPDC imager the quantum correlation imparts an additional factor of $2/\sqrt{3} + 1/4\sqrt{\pi}\mathcal{I}$.

To get a better understanding of the SNR behavior for pseudothermal and SPDC ghost imagers, we will consider three limiting cases. The first is the long integration-time limit in which $\Delta^2\text{Path}$ dominates the noise denominators for both the pseudothermal and SPDC ghost imagers, leading to the \em identical\/\rm\ maximum (saturation) SNR value
\begin{align}\label{eq:SNRsat}
\text{SNR}_{\rm sat} =\frac{e^{-4(\sigma_R^2+\sigma_S^2+\sigma_T^2)}}{\left[\Gamma+1\right]-e^{-4(\sigma_R^2+\sigma_S^2+\sigma_T^2)}},
\end{align}
regardless of the strength of the illumination. 

For our next two cases, we will assume the integration time is sufficiently short that SNR saturation is not approached.  Then, in the high-brightness ($\mathcal{I}\gg 1$) case, for which detection noise can be neglected, performance is dominated by source fluctuations, $\Delta^2\text{Source}$, yielding
\begin{align}\label{eq:SNRC_H}
\text{SNR}_{H,C} =\frac{T_I}{T_0}\frac{ \sqrt{2 \pi} \rho_L^2 e^{-4(\sigma_R^2+\sigma_S^2+\sigma_T^2)}}{A'_T (1+\beta^{-1})} \sT^2(\bm{\rho}_p),\\
\label{eq:SNRQ_H}
\text{SNR}_{H,Q} = \Omega_BT_I \sqrt{\frac{\pi}{8}}\frac{\rho_L^2 e^{-4(\sigma_R^2+\sigma_S^2+\sigma_T^2)}}{A'_T (1+\beta^{-1})} \sT^2(\bm{\rho}_p),
\end{align}
for pseudothermal ($C$) and SPDC ($Q$) imagers, respectively.  
With all parameters other than source coherence-time being equal, the SPDC high-brightness SNR exceeds that of the pseudothermal source by a factor $\Omega_BT_0 \gg 1$, where $T_0$ is the pseudothermal source's coherence time.  This is because the much faster fluctuations in the SPDC output fields lead to faster decorrelation, and hence higher SNR.  That said, however, the reader is reminded that cw SPDC operation is ordinarily in the low-brightness regime, hence we turn our attention now to that special case.  

In the low-brightness, $\mathcal{I}\ll 1$, regime, SNR behavior is dominated by the beat between the detection noises, $\Delta^2\text{Detect}$. Here we find that
\begin{align}\label{eq:SNRC_L}
&\text{SNR}_{L,C}=\frac{T_I}{T_0}\frac{16\sqrt{2}}{\sqrt{\pi}}\frac{A_p\eta^2\mathcal{I}^2}{\Omega_B T_0\rho_L^2}\sT(\bm{\rho}_p)\frac{A_b}{L^2}, \\
\label{eq:SNRQ_L}
&\text{SNR}_{L,Q}=\frac{T_I}{T_0}\frac{A_p\eta^2\mathcal{I}}{\pi \rho_L^2}\sT(\bm{\rho}_p)\frac{A_b}{L^2},
\end{align}
for the pseudothermal and SPDC cases.  
The SPDC SNR is linear in source brightness, whereas the pseudothermal SNR is quadratic in that parameter. This behavior is to be expected, because the SPDC low-brightness output approaches the biphoton state, which is known to have lower detection noise than classical-state light. Furthermore, $T_0$ for the pseudothermal source will greatly exceed $T_0$ for the SPDC, and, as noted in the preceding paragraph, $\Omega_BT_0 \gg 1$ holds for the pseudothermal source.  It follows that the low-brightness SPDC SNR is much higher than that of the pseudothermal imager.  The reader is cautioned to remember, however, that pseudothermal ghost imaging is typically performed in the high-brightness regime.  

It is also instructive to compare our SNR results---which apply to reflective ghost imaging through atmospheric turbulence---to the transmissive ghost imaging, no-turbulence results from \cite{Erkmen2009SNR}.  The principal differences between the SNR behaviors in these two cases is as follows.  Our SNR expressions have the average intensity reflection coefficient $\sT(\bp_p)$ appearing in lieu of the magnitude-squared of the field reflection coefficient $|T(\bp_p)|^2$ that is seen in the transmissive case.  Moreover, the transmissive-case SNR from \cite{Erkmen2009SNR} has neither target-induced speckle nor turbulence contributions, so it has neither $\Delta^2\text{Path}$ term nor any turbulence-induced noise-magnification factors.  The reflective case's $\Delta^2\text{Source}$ term has a factor of  $(1+\beta^{-1})$ and its $\Delta^2\text{Mix}$ term has a factor of $(4/3+\beta^{-1})$, but both $\beta^{-1}$ contributions  are absent from the transmissive case. These additional reflective-case contributions are direct results of the rough-surface scattering.  In both transmissive and reflective operation there is averaging of the source randomness at the target, but in the reflective case the fields scatter incoherently off the target, and their Fourier components are averaged at the bucket detector. This second averaging depends on the ratio of the source size to the bucket size, i.e., $\beta$. As $\beta$ increases, the bucket detector captures more light, this term becomes less significant, and we approach the transmissive behavior for these terms.  Finally, the reflective-SNR terms associated with the bucket detector measurement in $\Delta^2\text{Detect}$ and $\Delta^2\text{Mix}$ are scaled by $L^2/A_b$, relative to the corresponding terms in the transmissive case.  This scale factor is the inverse of the solid angle subtended at the target by the bucket detector.  Its presence in the reflective case is a consequence of the quasi-Lambertian scattering from the rough surface

It follows from the SNR differences highlighted above that the low-brightness SNR asymptotes for transmissive and reflective operation only differ by the latter's including the solid-angle subtense factor $A_b/L^2$. Their high-brightness SNR asymptotes are also similar, with the only difference being the reflective case's factor of $(1+\beta^{-1})$. The most profound difference, however, is due to the $\Delta^2\text{Path}$ term that is only present for reflective operation.  This term implies that the SNR of reflective ghost imaging has a maximum value, SNR$_{\rm sat}$, whereas no such saturation occurs in transmissive ghost imaging, i.e., its SNR can grow without bound as $T_I/T_0$ increases.  Furthermore, in reflective imaging, this limiting effect can be quite severe: even with no turbulence, for $\beta=1$ we find ${\rm SNR}_{\rm sat} =3.26$, and for $\beta = 2$ we have ${\rm SNR}_{\rm sat}=5.54$. So, for realistic standoff sensing, the SNR will be limited to single-digit values if no further measures are taken to suppress the fluctuations arising from target-induced speckle and atmospheric turbulence.

\section{Computational Ghost Imager}\label{sec:Comp}
We have developed the performance of reflective ghost imaging from classical and nonclassical sources in a quantum Gaussian-state framework. Next we consider a computational ghost imager, a variant of the pseudothermal case in which the ``chaotic'' signal field is created by deterministically modulating the wavefront of a laser with a spatial light modulator (SLM), and no reference-arm light beam is required.  Instead, knowledge of the SLM phase pattern is used to calculate the signal-beam intensity at the target for use in lieu of a reference-beam measurement \cite{Shapiro2008Comp}. If an independent pseudorandom phase process is applied to each SLM pixel, then far-field propagation can be argued to yield Gaussian field statistics per the Central Limit Theorem.  Thus it is appropriate for us to model the computational ghost imager's far-field statistics with a Gaussian-Schell model, taking the source's coherence length to be approximately the width of an SLM pixel, and source's intensity diameter to be the diameter of the SLM. 

The laser that illuminates the SLM will be taken to emit coherent-state light, hence the SLM's output beam is in classical state and semiclassical theory will give quantitatively identical results for its far-field photodetection statistics as the full quantum treatment \cite{Erkmen2008Unified, Erkmen2010AOP,Shapiro2008Comp}.  Also, because the reference field at the target is computed, rather than measured, it is most naturally modeled as a complex-valued Gaussian random process.  Consequently, and without loss of generality, we choose to use semiclassical theory for our analysis of computational ghost imaging because it lets us deal with only classical random processes, rather than a combination of quantum field operators and classical random processes. We thus use $E_S'(\bp,t)$ to represent the complex envelope of the target-plane signal field, and $\tilde{E}_R'(\bp,t)$ to represent the computed complex-envelope of the field that would have been present on the CCD camera.  These two complex-valued, zero-mean, Gaussian random processes are completely characterized by their nonzero autocorrelation and cross-correlation functions, which we take to be as given in Eqs.~\eqref{eq:AutoCorr}--\eqref{eq:CCorrPIS} for their field-operator pseudothermal counterparts.  That these correlation functions are consistent with classical random-process theory follows from the pseudothermal source's producing classical-state light.  

The principal distinction between computational and pseudothermal ghost imaging---which is also its main advantage---is that there is neither detection noise nor turbulence on its computed reference-arm intensity, which is found via
\begin{align}
\tilde{i}_p(t) = q\eta\int d\tau\, \tilde{P}_p(\tau) h(t-\tau), 
\end{align}
with $\tilde{P}_p(t) = A_p|\tilde{E}'_R(\bp_p,t)|^2$. The bucket detector photocurrent, in semiclassical theory, is given by
\begin{align}
i_b(t) = \int d\tau\, [q\eta P_b(\tau) + \Delta i_b(\tau)] h(t-\tau), 
\end{align}
where $P_b(t)=\int_{{\cal A}_b} \!d\bp\,|E_T'(\bp,t)|^2$ is the impinging photon flux, and conditioned on that photon flux the shot noise $\Delta i_b(t)$ is a zero-mean random process with autocorrelation function $\avg{\Delta i_b(t_1) \Delta i_b(t_2)} = q^2 \eta P_b(t_1)\delta(t_1-t_2)$. The computational ghost image is then obtained from the correlation function
\begin{align}
\label{CGI}
\tilde{C}(\bp_p) = \frac{1}{T_I}\int_{-T_I/2}^{T_I/2}\!dt\, \tilde{i}_p(t) i_b(t).
\end{align}

Because the shot-noise is zero mean, given $P_b(t)$, it does not contribute to $\langle \tilde{C}(\bp_p)\rangle$, hence the computational ghost imager's spatial resolution and image contrast analysis is identical to the Sec.~\ref{sec:ResCon} treatment of the pseudothermal ghost imager once we substitute the computational system's classical random processes for the pseudothermal imager's field operators, and then let $\rho_R\rightarrow \infty$, and set $\sigma_R^2=0$ to account for the absence of turbulence on the computed reference. Using this approach we immediately find the spatial resolution to be $\rho_L\sqrt{\tilde{\alpha}}$ where
\begin{align}
\tilde{\alpha} = \frac{2\rho_S^2+a_0^2}{2 \rho_S^2},
\end{align}
and the image contrast is identical to that given for the pseudothermal imager in Eq.~\eqref{eq:CContrast}.  In the latter regard we note that we could replace $\tilde{i}_p(t)$ in Eq.~(\ref{CGI}) with the zero-mean process, $\Delta\tilde{i}_p(t) \equiv \tilde{i}_p(t) -\langle \tilde{i}_p(t)\rangle$, thereby achieving the same high-contrast operation as is realized with {\sc ac}-coupling in pseudothermal ghost imaging.

Turning now to the computational ghost imager's signal-to-noise ratio, we find more significant differences between its analysis and that of the pseudothermal ghost imager.  In our Sec.~\ref{sec:SNR} SNR analysis, the second moment of the photocurrent cross-correlation function involved an eighth-order field moment, two sixth-order field moments, a fourth-order field-moment, three fourth-order moments of the turbulence, and a fourth-order moment of the target's field-reflection coefficient.  For the computational ghost imager's SNR analysis we only have to evaluate an eighth-order field moment, a sixth-order field moment, two fourth-order moments of the turbulence, and the fourth-order moment of the target's field-reflection coefficient. This simplification is because the other terms involved the path and detection noises on the reference arm, which are absent from the computational configuration. The computational SNR is then given by
\begin{widetext}
\begin{align}\label{eq:CompSNR}
\text{SNR}_{\tilde{C}}=
\frac{\displaystyle \sT^2(\bp_p)\frac{T_I}{T_0}e^{-4(\sigma_S^2+\sigma_T^2)}}
{\displaystyle\frac{A'_T(1+\beta^{-1})}{\sqrt{2\pi}\rho_L^2}  
+\sT^2(\bp_p)\frac{T_I}{T_0}(\left[\Gamma+1\right]-e^{-4(\sigma_S^2+\sigma_T^2)})
+\frac{\sT(\bp_p)}{\eta\mathcal{I}}\frac{ L^2}{A_b}}.
\end{align}
\end{widetext}
It follows that computational ghost imaging shares the same saturation SNR as pseudothermal and SPDC ghost imaging, from Eq.~(\ref{eq:SNRsat}), with $\sigma_R^2$ set to zero because there is no turbulence in the computational imager's reference arm.  Likewise, computational ghost imaging has the same high-brightness SNR asymptote as pseudothermal ghost imaging, from Eq.~(\ref{eq:SNRC_H}), with $\sigma_R^2 = 0$.  
On the other hand, computational ghost imaging has a significantly improved low-brightness SNR asymptote, given by
\begin{align}\label{eq:SNRComp_L}
\text{SNR}_{\tilde{C},L} = \frac{T_I}{T_0}\eta\mathcal{I}\sT(\bp_p)\frac{A_b}{L^2},
\end{align}
in comparison with \em both\/\rm\ pseudothermal and SPDC ghost imaging.  

For quantitative comparison between the SNRs of the pseudothermal, SPDC, and computational systems we will first have to put them on equal footing. The dimensionless brightness quantity $\mathcal{I}$ that appears in our SNR formulas is the photon flux per spatiotemporal mode of the source.  However, because of the dramatically different coherence times of the SPDC ($\Omega_BT_0 \ll 1$) and the pseudothermal and SLM sources ($\Omega_BT_0 \gg 1$), comparisons based on equal $\mathcal{I}$ values will be for sources with dramatically different photon fluxes, whereas we would prefer to compare SNRs at equal photon fluxes.  To do so we introduce the dimensionless quantity $\mathcal{I}_\Omega = \mathcal{I}\Omega_B T_0 = P \rho_0^2/a_0^2\Omega_B$, which is the photon flux per source spatial mode and detector temporal mode. Because all three systems are taken to have detectors with identical characteristics, identical values for $\mathcal{I}_\Omega$ will yield identical photon fluxes. Making this substitution we can easily compare the low-brightness SNRs from Eqs.~\eqref{eq:SNRC_L}, \eqref{eq:SNRQ_L}, and \eqref{eq:SNRComp_L}. In the extreme low-brightness regime, the computational imager has the best SNR, followed by the SPDC and pseudothermal imagers. We can also compare the high-brightness limits Eqs.~\eqref{eq:SNRC_H} and \eqref{eq:SNRQ_H}, where the shorter coherence time of the SPDC source gives it a clear advantage.

Because the different sources exit their low- and high-brightness SNR regimes and enter into SNR saturation for different photon fluxes, assessing that behavior requires us to explicitly choose some operating parameters. First, we assume the operating wavelength is $\lambda = 1.5\, \mu$m, the source coherence length is $\rho_0= 0.15/\pi\,$mm, the source intensity radius is $a_0= 3\,$cm, and that the bucket detector matches the source size, i.e., $A_b = \pi a_0^2$. These give us the on-target parameters of $\rho_L = 0.05/\pi\,$m, and $a_L = 10 \,$m. The photodetectors are taken to have $\eta=0.9$ quantum efficiency, with the CCD pixel area being $A_p = 0.1 \rho_L^2$.  We assume the target is at $L= 1\,$km range, that it has an effective area of $A'_T = 50\, \text{m}^2$, and $\sT(\bp_p) = 1$ at the point for which we will perform our SNR evaluations. All paths are assumed to have moderate-strength ($C_n^2 = 10^{-14}\,{\rm m}^{2/3}$) turbulence. Finally, we take the pseudothermal and computational sources to have temporal coherence times satisfying $T_0 = 10^3/\Omega_B \,$Hz, while for the SPDC source $T_0 =1/ 10^3\Omega_B \,$Hz.

\begin{figure}
\includegraphics[width=3.5in]{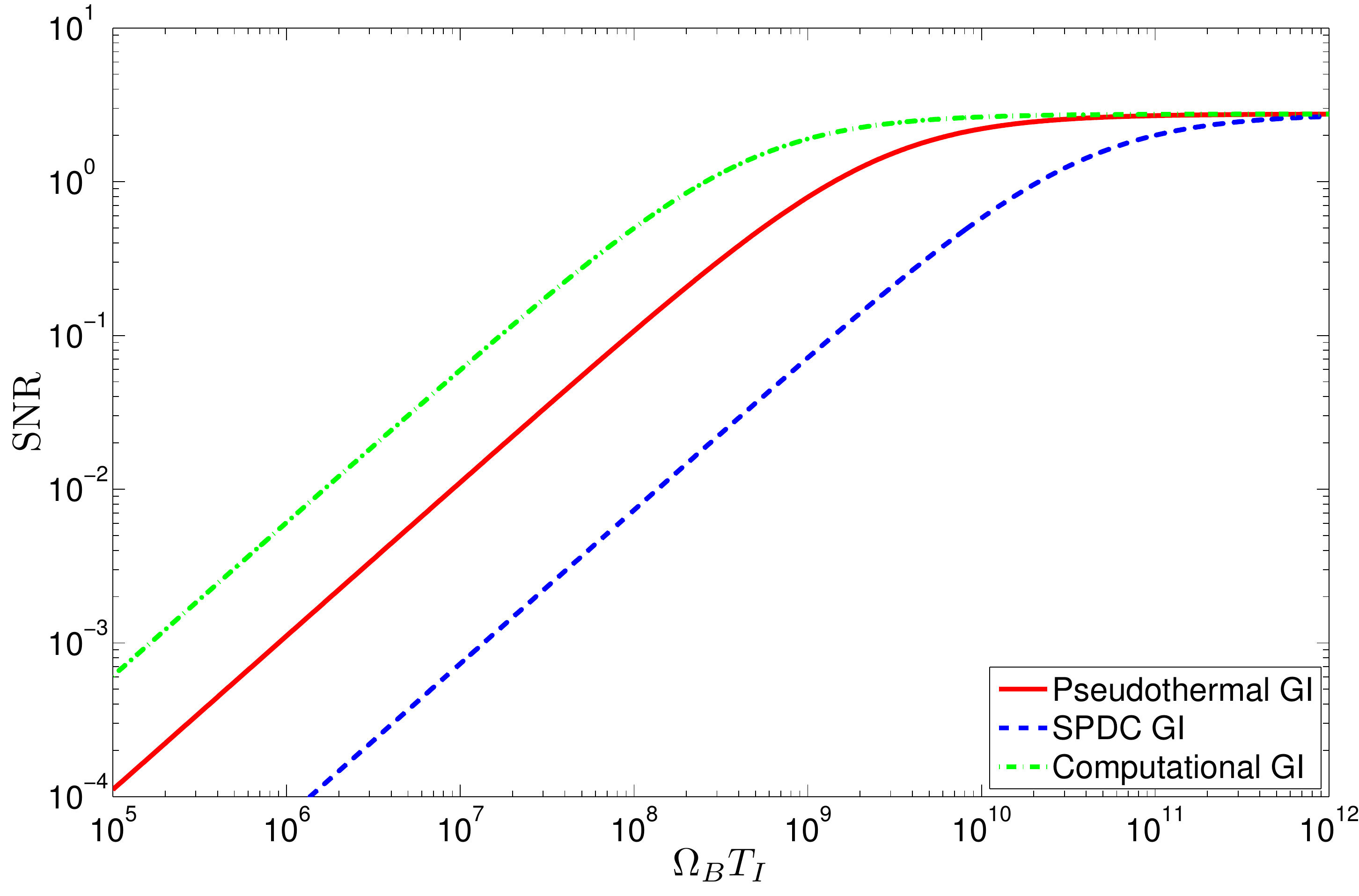}
\caption{(Color online) SNRs for pseudothermal, SPDC, and computational ghost imaging vs.\ normalized integration time $\Omega_B T_I$.  The source brightness is $\mathcal{I}_\Omega = 1$, so that the SPDC output is a stream of biphotons and all three systems are well below their respective high-brightness SNR asymptotes.  Other parameters used in computing these curves are given in the text.}
\label{fig:SNR_L}
\end{figure}

\begin{figure}
\includegraphics[width=3.5in]{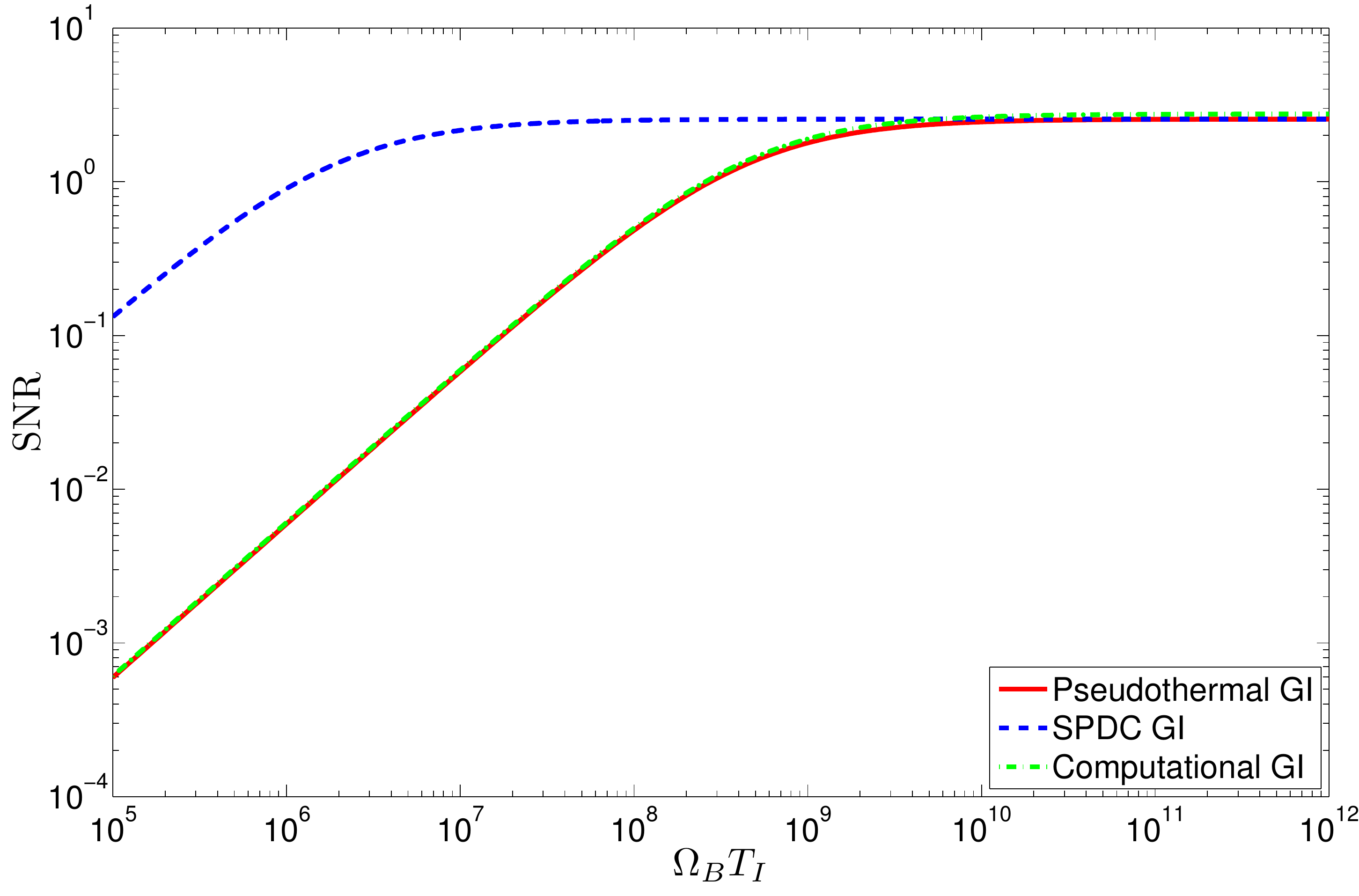}
\caption{(Color online) SNRs for pseudothermal, SPDC, and computational ghost imaging vs.\  normalized integration time $\Omega_B T_I$.  The source brightness is $\mathcal{I}_\Omega = 10^4$, so that the SPDC output appears to be classical and the pseudothermal and computational ghost imagers are in their high-brightness limits wherein they only differ by a factor of $e^{4\sigma_R^2}$.  Other parameters used in computing these curves are given in the text.}
\label{fig:SNR_H}
\end{figure}

We consider two scenarios:   first, when the source is weak enough ($\mathcal{I}_\Omega = 1$) that the SPDC output is still a stream of biphotons, but we are not deep into the low-brightness regime; and second, when the source is sufficiently strong ($\mathcal{I}_\Omega = 10^4$) to make the SPDC output appear classical \cite{footnote5}. The resulting SNR curves for these two cases are shown in Figs. \ref{fig:SNR_L} and \ref{fig:SNR_H}, respectively.  From Fig.~\ref{fig:SNR_L} we see that the SPDC imager is the worst performer when it is in its biphoton regime, while the computational ghost imager is to the top performer there.  From Fig.~\ref{fig:SNR_H}, however, we see that when the SPDC output appears to be classical, its much shorter coherence time allows it to reach the saturation-SNR limit well before the other two systems, whose SNR curves are nearly identical.

\section{Discussion}\label{sec:Discussion}
We have derived the key performance characteristics of reflective ghost imaging in the context of realistic standoff sensing applications. This was done within a Gaussian-state framework that allowed for analysis of classical pseudothermal sources and the nonclassical output of an SPDC. Our results for the spatial resolution, image contrast, and SNR were analogous to those calculated for transmissive ghost imaging in the absence of turbulence \cite{Erkmen2008Unified, Erkmen2009SNR}, with a few significant differences that result from the roughness of the target surface, turbulence on the optical paths, and the standoff nature of the measurement. First we shall discuss the impact of rough-surface scattering in the absence of turbulence, then we shall discuss the additional effects incurred when turbulence is present.  

Rough-surfaced targets scatter their impinging illumination in a random manner, casting random speckle patterns in their far fields that, on average, correspond to a quasi-Lambertian distribution.   Consequently, the ghost-imager's bucket detector will, on average, collect only a fraction, $A_b/L^2$, of the target-scattered light.  This angular subtense factor---which is absent in transmissive ghost imaging---appears in the average photocurrent cross-correlation functions in Eqs.~\eqref{eq:CAvgMean} and \eqref{eq:QAvgMean}, and the noise terms associated with the bucket detector, i.e., $\Delta^2\text{Detect}$ and $\Delta^2\text{Mix}$. The surface roughness has another major effect on the SNR:  its time-independent speckle pattern implies the existence of a finite saturation-SNR value in the limit of long integration time.  

The target-speckle contribution that leads to a saturation SNR appears in $\Delta^2\text{Path}$, and is a function of $\beta$, the ratio of the source size to the bucket detector size. Here we may incur a trade-off between spatial resolution and SNR. Specifically, by enlarging the bucket detector we can increase the averaging of the target speckle, and thus the saturation SNR, without affecting the imager's spatial resolution. Ultimately, however, there will be a practical limit beyond which the receiving aperture size can no longer be increased. Beyond that point $\beta$, and thus the saturation SNR, can only be increased by decreasing the source size $a_0$.  However, as seen in Eq.~\eqref{eq:Res}, spatial resolution is inversely proportional to $a_0$, so increasing the saturation SNR in this manner will degrade spatial resolution.

The effects of turbulence on the image contrast and spatial resolution of reflective ghost imaging are similar those previously found for transmissive ghost imaging \cite{Cheng2009}.  In particular:  turbulence does not change the image contrast;  turbulence in the target-to-bucket path has no effect on spatial resolution; and turbulence on the signal and reference paths degrades spatial resolution in the same manner, i.e., degradation occurs when they become smaller than the source size. In both SPDC and pseudothermal ghost imaging it is therefore advantageous to propagate the reference field through a controlled (turbulence-free) environment.  Turbulence also degrades ghost-image SNR in two ways: first, its logamplitude fluctuations magnify several existing noise terms; and second, these fluctuations also contribute to the time-independent noise term $\Delta^2\text{Path}$, decreasing the saturation SNR from its speckle-only value.

In computational ghost imaging, the reference field is computed, removing the turbulence on that path, and thus improving both the spatial resolution and the SNR. The computational case also has significantly reduced noise for low-brightness illumination, as there is no longer detection noise on the reference arm, making its low-brightness SNR behavior superior to that of both pseudothermal and SPDC ghost imaging.   More generally, Figs.~\ref{fig:SNR_L} and \ref{fig:SNR_H} show that the computational ghost imager has a higher SNR than the pseudothermal system in both low-brightness and high-brightness operation, and than the SPDC system in its low-brightness (biphoton) regime. However, in high-brightness operation, the SPDC ghost imager's significantly shorter coherence time enables it to reach its saturation SNR much quicker than either the computational or the pseudothermal ghost imagers.  Although a cw SPDC will not be able to reach that high-brightness regime, pulsed systems are capable of doing so \cite{Gouet2009}.

Insofar as standoff sensing is concerned, the essential conclusion to be drawn from the preceding summary of our work is simple:  the computational approach is the ghost-imaging configuration that shows the most promise for this application.  It has the best spatial resolution, the highest saturation-SNR, and general SNR performance second only to high-brightness SPDC operation.  In addition, the computational approach obviates the need for a physical reference path, which makes it naturally amenable to obtaining 3D ghost images.  In particular, pseudothermal and SPDC ghost imaging require that reference-arm measurements be made on an intensity pattern corresponding to the one that is projected onto the target.  As shown theoretically in \cite{Shapiro2008Comp} and experimentally in \cite{Bromberg2009}, computational ghost imaging allows reference intensity patterns to be computed at a variety of target ranges so that ghost images can be formed for these target ranges from the same bucket-detector data.  This range sectioning is something that \em cannot\/\rm\ be done, without separate measurements for each possible target range, in pseudothermal or SPDC ghost imaging. 

We have reserved our final comment for our choice of image reconstruction via photocurrent cross-correlation.   It has already been demonstrated \cite{Bromberg2009,Katz2009} that ghost imaging reconstruction can be performed in transmission through more advanced methods, such as compressive sensing. A preliminary no-turbulence performance comparison between a simple cw-laser radar and ghost imaging in reflection via cross correlation has shown their similarity in spatial resolution and SNR \cite{Hardy2010}.   If advanced image-reconstruction techniques can be successfully applied to computational ghost imaging, the resulting system could have notable advantages over current laser radars.

\begin{acknowledgments}
This work was supported by the U.S. Army Research Office Multidisciplinary University Research Initiative Grant No.\ W911NF-05-1-0197, the W. M. Keck Foundation Center for Extreme Quantum Information Theory, the DARPA Information in a Photon Program under U.S. Army Research Office Grant No.\ W911NF-10-1-0404, and the DARPA Quantum Sensors Program under AFRL Contract No.\ FA8750-09-C-0194,  

\end{acknowledgments}


\begin{thebibliography}{99}
\bibitem{footnote1} For the sake of generality, we have allowed there to be turbulence in the reference path, although a lens could be used to cast a minified reference-beam pattern onto a CCD camera within a protected environment near the source, hence avoiding turbulence on that path.
\bibitem{footnote3} A similar square-law approximation is available for cases in which the turbulence strength is not uniformly distributed along each propagation path, but we have opted to limit our attention to uniform distributions.
\bibitem{footnote4} In quantum photodetection, the currents at the outputs of the bucket and CCD detectors are classical stochastic processes, $i_b(t)$ and $i_p(t)$, whose statistics are equivalent to those of the quantum measurements $\hat{i}_b(t)$ and $\hat{i}_p(t)$ given in the next paragraph.
\end{thebibliography}
\end{document}